\documentclass[lettersize,journal]{IEEEtran}
\usepackage{amsmath,amsfonts}
\usepackage[ruled,lined,linesnumbered]{algorithm2e}
\usepackage{array}
\usepackage{color}
\usepackage{textcomp}
\usepackage{stfloats}
\usepackage{url}
\usepackage{verbatim}
\usepackage{graphicx}
\usepackage{cite}
\usepackage[colorlinks, citecolor=black]{hyperref}

\usepackage{mathrsfs}
\usepackage{multirow,diagbox}
\usepackage{verbatim}
\usepackage{color}
\usepackage{colortbl}
\usepackage{makecell}
\usepackage{cellspace}
\usepackage{stfloats}
\usepackage{longtable}
\usepackage{booktabs}
\usepackage{array,multirow,multicol,graphicx}
\usepackage{color}
\usepackage{float}
\usepackage{makecell}
\usepackage[x11names,dvipsnames,table]{xcolor} 
\usepackage{colortbl}
\usepackage{multirow}
\usepackage[font=footnotesize]{caption}
\usepackage{subcaption}
\usepackage{lipsum}
\definecolor{mygray}{gray}{0.6}
\definecolor{myblue}{rgb}{0.8,0.85,1}
\usepackage{array}
\newcolumntype{L}[1]{>{\raggedright\let\newline\\\arraybackslash\hspace{0pt}}m{#1}}
\newcolumntype{C}[1]{>{\centering\let\newline\\\arraybackslash\hspace{0pt}}m{#1}}
\newcolumntype{R}[1]{>{\raggedleft\let\newline\\\arraybackslash\hspace{0pt}}m{#1}}

\usepackage{array, tabularx, boldline}
\usepackage{eurosym}
\usepackage{amstext} 
\usepackage{amssymb} 
\usepackage{tabularray} 
\newcommand{\smallsix}[1]{\fontsize{6.6pt}{2pt}\selectfont #1\normalsize}

\definecolor{mygray}{gray}{0.9} 

\DeclareRobustCommand{\officialeuro}{%
  \ifmmode\expandafter\text\fi
  {\fontencoding{U}\fontfamily{eurosym}\selectfont e}}
\usepackage{amsthm}
\theoremstyle{plain}

\begin{document}

\title{Threats, Attacks, and Defenses in Machine Unlearning: A Survey}


\author{Ziyao Liu, Huanyi Ye, Chen Chen, Yongsen Zheng and Kwok-Yan Lam

%
%
\thanks{Ziyao Liu, Chen Chen, Huanyi Ye, Yongsen Zheng and Kwok-Yan Lam are with Nanyang Technological University, Singapore.  
E-mail: liuziyao@ntu.edu.sg,  huanyi001@e.ntu.edu.sg, chen.chen@ntu.edu.sg, yongsen.zheng@ntu.edu.sg,  kwokyan.lam@ntu.edu.sg.}
}

\markboth{Journal of \LaTeX\ Class Files,~Vol.~14, No.~8, November~2023}%
{Shell \MakeLowercase{\textit{et al.}}: A Sample Article Using IEEEtran.cls for IEEE Journals}

\IEEEpubid{0000--0000/00\$00.00~\copyright~2021 IEEE}

\maketitle
\IEEEpubidadjcol

\begin{abstract}
Machine Unlearning (MU) has recently gained considerable attention due to its potential to achieve Safe AI by removing the influence of specific data from trained Machine Learning (ML) models. This process, known as knowledge removal, addresses AI governance concerns of training data such as quality, sensitivity, copyright restrictions, and obsolescence. This capability is also crucial for ensuring compliance with privacy regulations such as the Right To Be Forgotten (RTBF). Furthermore, effective knowledge removal mitigates the risk of harmful outcomes, safeguarding against biases, misinformation, and unauthorized data exploitation, thereby enhancing the safe and responsible use of AI systems. Efforts have been made to design efficient unlearning approaches, with MU services being examined for integration with existing machine learning as a service (MLaaS), allowing users to submit requests to remove specific data from the training corpus. However, recent research highlights vulnerabilities in machine unlearning systems, such as information leakage and malicious unlearning, that can lead to significant security and privacy concerns. Moreover, extensive research indicates that unlearning methods and prevalent attacks fulfill diverse roles within MU systems. This underscores the intricate relationship and complex interplay among these mechanisms in maintaining system functionality and safety. This survey aims to fill the gap between the extensive number of studies on threats, attacks, and defenses in machine unlearning and the absence of a comprehensive review that categorizes their taxonomy, methods, and solutions, thus offering valuable insights for future research directions and practical implementations.

\end{abstract}

\begin{IEEEkeywords}
Machine unlearning, MLaaS, threats, attacks, defenses
\end{IEEEkeywords}

\section{Introduction}
\label{sec:introduction}
In recent years, the concept of knowledge removal in AI systems has garnered significant attention from both academia and industry, primarily due to security, privacy, and safety concerns. There are some examples as follows:
\begin{itemize}
    \item Some training data may be considered sensitive, necessitating its removal from the model to prevent the leakage of sensitive information.
    \item In fields like healthcare, navigation, or forensics, outdated or incorrect training data must be removed from the model to maintain its safety and reliability.
    \item Training data protected by copyright requires the removal of knowledge gained from them from the trained machine learning (ML) model. This step is crucial to comply with commercial regulations and avoid conflicts before the model's deployment or release.
    \item Additionally, data privacy regulations such as GDPR \cite{regulation2018general}, APPI \cite{iwase2019overview}, and CCPA \cite{goldman2020introduction} emphasize data rights, including the Right To Be Forgotten (RTBF), allowing individuals to request the deletion of their personal data. This underscores the importance of effective knowledge removal to comply with these regulations.
\end{itemize}

\IEEEpubidadjcol

\begin{figure}[t]
\centering
		\centering
\includegraphics[width=0.95\linewidth]{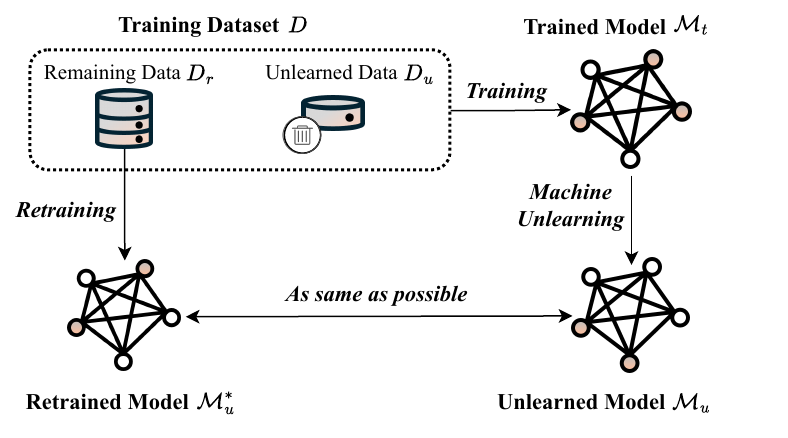}
		\caption{An illustration of machine unlearning.}
		\label{fig:mu_scheme}
\end{figure}

In AI systems, removing knowledge requires the deletion of specific data and its effects from a trained ML model. As such, machine unlearning (MU) \cite{xuh2023machine} emerged as a crucial enabler of this process. As depicted in Figure \ref{fig:mu_scheme}, an initial, yet naive, approach to implementing MU is through retraining. This method involves discarding the current model and retraining it from scratch with the remaining data after the removal of the data to be unlearned. Nevertheless, this approach can incur substantial computational expenses, making it impractical for scenarios involving large-scale models or extensive datasets. In response to this challenge, a range of efficient MU methodologies have recently been developed, allowing the unlearned model to be obtained from the existing trained model without the need for retraining from scratch \cite{cao2015towards,warnecke2021machine,zhao2023federated,liu2022right,wang2024server,hu2024learn,nguyen2022survey,xu2023machine,qu2023learn,lin2024blockchain,qiu2023fedcio,ding2023strategic,shao2024federated,ding2023incentivized,shaik2023exploring,ding2023incentive,lin2024scalable}.

\begin{figure*}[t]
    \centering
    \includegraphics[width=0.9\linewidth]{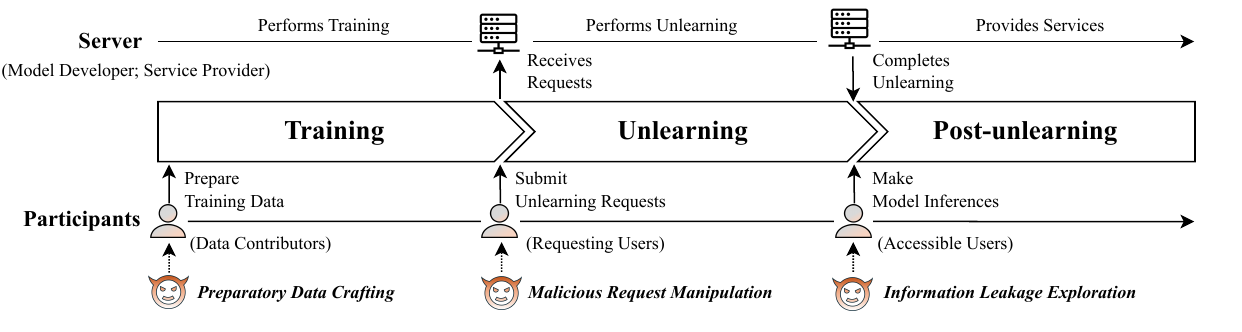}
    \caption{The overall workflow for machine unlearning systems.}
    \label{fig:workflow}
\end{figure*}

Furthermore, with the widespread adoption of machine learning as a service (MLaaS), AI service providers are working to integrate machine unlearning functionality into their offerings to meet the growing emphasis on the RTBF in privacy regulations and the increasing need for knowledge removal from AI models. However, research indicates that vulnerabilities are present in MLaaS systems that provide unlearning services. As illustrated in Figure \ref{fig:workflow}, adversarial users may launch various attacks at different stages of the MLaaS process with unlearning functionality, leading to significant security issues and privacy concerns (see detailed discussion in subsequent sections). These threats pose significant challenges to the safe deployment of machine unlearning systems.

The increasing emphasis on security and privacy in unlearning systems has led to a clearer recognition of their vulnerabilities and the threats they face, as revealed by related studies. Moreover, the complex roles of unlearning methods and the prevalence of attacks have become more evident, underscoring their intricate interaction in maintaining the functionality and safety of MU systems. Given this backdrop, a thorough examination of existing research on threats, attacks, and defenses in MU systems is indispensable. Such a comprehensive review would not only categorize their taxonomy, methods, and solutions but also provide valuable insights for future research directions and practical implementations. Thus, this survey is developed to address this existing gap, aiming to offer a holistic overview and critical analysis of the field.

\noindent\textbf{Comparison with other surveys.}
Recently, several studies begin to study and analyze machine unlearning methodologies, resulting in a clear categorization of the existing survey research. These categories include general surveys on machine unlearning, highlighted by works such as \cite{nguyen2022survey,xuh2023machine,xu2023machine,shaik2023exploring,li2024machine,qu2023learn,liu2024survey,wang2024machine}, which cover a wide range of topics from basic concepts and methods to the challenges faced in the field. Additionally, specific attention has been given to federated unlearning (FU) through studies like \cite{liu2023survey,jeong2024sok,romandini2024federated,yang2023survey}. These surveys focus on the unique issues and uses of federated unlearning compared to traditional unlearning approaches. Furthermore, potential risks and threats within federated unlearning are examined in \cite{wang2023federated}. Benchmarks and surveys on MU approaches over large language models (LLMs) are given in \cite{li2024wmdp,lynch2024eight,thaker2024guardrail,si2023knowledge,liu2024rethinking}.

\begin{itemize}
    \item General surveys on MU: \cite{nguyen2022survey,xuh2023machine,xu2023machine,shaik2023exploring,qu2023learn,li2024machine,liu2024survey,wang2024machine}
    \item Surveys on FU: \cite{liu2023survey,jeong2024sok,romandini2024federated,yang2023survey,wang2023federated}
    \item Benchmarks and surveys on MU for LLMs: \cite{li2024wmdp,lynch2024eight,thaker2024guardrail,liu2024rethinking,si2023knowledge,liu2024machine}
\end{itemize}

To the best of our knowledge, no comprehensive survey currently tackles security and privacy issues within the context of machine unlearning, particularly focusing on threats, attacks, and defensive strategies. This gap in the literature strongly motivates our work in compiling this survey. Our focus on categorizing and scrutinizing these elements aims to illuminate strategies for protecting machine unlearning systems, thereby enhancing their robustness and safety.

\noindent\textbf{Summary of contributions.} The main contributions of this survey are listed as follows:
\begin{enumerate}
    \item  We present a unified workflow for machine unlearning, and drawing from this framework as well as threat models, we propose a novel taxonomy of threats, attacks, and defenses within MU systems.
    \item We carry out a detailed analysis of the current threats, attacks, and defenses in machine unlearning, along with an examination of the intricate relationships among unlearning methods, attacks, and defensive strategies.
    \item We provide an in-depth discussion concerning potential vulnerabilities and threats in unlearning systems and the challenges in defenses against these threats and attacks, highlighting key areas for future research exploration.
\end{enumerate}

\noindent\textbf{Organisation of the paper.} The remainder of this paper is structured as follows: Section \ref{sec:preliminaries} outlines the key concepts fundamental to this survey. Section \ref{sec:workflow} introduces a unified MU workflow to support the analysis and taxonomy presented in the subsequent sections. Section \ref{sec:overview} provides an overview of the threats in MU systems and the various roles of attacks and defenses. Section \ref{sec:threat_models} provides the threat models for machine unlearning systems. Building on this foundation, we conduct a thorough review of threats within the MU workflow, complete with a taxonomy, in Section \ref{sec:threats_in_unlearning}. Section \ref{sec:unlearning_as_defenses} discusses the role of unlearning in defense mechanisms, while Section \ref{sec:attacks_for_evaluation} examines how attacks can serve as a means for unlearning evaluation, followed by a summary in Section \ref{sec:summary}. Section \ref{sec:discussions_and_promising_directions} delves into challenges and potential directions for future research. Finally, Section \ref{sec:conclusions} provides a summary and concludes the paper.

\section{Preliminaries}
\label{sec:preliminaries}

In this section, we provide the foundational concepts for understanding the topics covered in this survey. We introduce machine unlearning, followed by an overview of machine learning as a service. We also discuss key attacks on AI models, including poisoning attacks and membership inference attacks, to lay the groundwork for the threat and defense analysis in later sections.

\subsection{Machine Unlearning}
In machine unlearning, as illustrated in Figure \ref{fig:mu_scheme}, the training dataset $D$ comprises two subsets, namely $D_u$ and $D_r$, representing the data points to be unlearned and the remaining data points, respectively, wherein $D_r = D \backslash D_u$. Denote $\mathcal{M}_t=\mathcal{M}(D)$ as the model trained on the entire dataset $D$. The goal of MU is to produce an unlearned model $\mathcal{M}_u$ that closely resembles the model $\mathcal{M}_u^*$ trained solely on $D_r$, i.e., $\mathcal{M}_u^*=\mathcal{M}(D_r)$.

Machine unlearning can be classified as either exact unlearning or approximate unlearning \cite{xuh2023machine}. Exact unlearning techniques typically involve retraining but limit the scope of data involved in order to enhance efficiency compared to naive retraining approaches. On the other hand, approximate unlearning adjusts the parameters of the existing trained model to create an unlearned model that approximates one that would be obtained through retraining from scratch. This approach often requires tradeoffs between unlearning performance and efficiency. Interested readers are referred to the relevant survey papers on machine unlearning, such as \cite{xuh2023machine,xu2023machine,qu2023learn,liu2023survey,yang2023survey}, for more comprehensive details about MU methodologies and taxonomy.

\subsection{Machine Learning as a Service}

Machine Learning as a Service (MLaaS) is a cloud-based service model that provides machine learning tools and infrastructure as part of a broader service offering. In an MLaaS framework, service providers host machine learning models and processing power, allowing users to submit data and receive predictions or insights without needing to develop, train, or maintain their own models \cite{ribeiro2015mlaas}. As a result, MLaaS has gained widespread adoption due to its ease of use, scalability, and accessibility, enabling organizations to focus on data and application logic rather than model development and infrastructure management. Current research efforts on MLaaS include but are not limited to privacy-preserving MLaaS \cite{hesamifard2018privacy,liu2020mpc} that ensure user data privacy during model training and prediction, scalability and efficiency to handle large-scale data and complex models with improved computational performance and resource management \cite{weng2022mlaas,zhang2020mlmodelci}, security and adversarial robustness to secure MLaaS systems against adversarial attacks like data poisoning and model evasion, enhancing system robustness. \cite{yu2020cloudleak,kesarwani2018model}.

\subsection{Attacks on AI Models}

Attacks on AI models refer to intentional actions taken by adversaries to manipulate, exploit, or undermine the performance and integrity of machine learning systems. These attacks can occur at various stages of the AI lifecycle, including data collection, training, and deployment. Common types of attacks include adversarial attacks \cite{chakraborty2018adversarial}, where small, carefully crafted changes are made to input data to deceive the model into making incorrect predictions, and data poisoning attacks \cite{tian2022comprehensive}, where malicious data is introduced into the training set to degrade the model’s performance. Other threats, such as model inversion \cite{fredrikson2015model} and membership inference \cite{shokri2017membership}, aim to extract sensitive information about the training data or individuals whose data was used to train the model. Since data poisoning attacks and membership inference attacks are widely adopted in the related works reviewed in this survey, we dedicate additional space to the preliminaries of these two types of attacks to facilitate a better understanding of their mechanisms in attacks on MU systems and the defenses discussed in the subsequent sections.

\textbf{Poisoning attack.} Poisoning attacks are a form of attack that aims to corrupt a machine learning model by injecting malicious data into its training set \cite{tian2022comprehensive}. These attacks are broadly categorized into untargeted and targeted poisoning attacks. Untargeted poisoning attacks aim to degrade the overall performance of the model by introducing enough corrupted data to lower its accuracy across a wide range of inputs. In contrast, in targeted poisoning attacks, the adversary focuses on manipulating the model to misclassify specific inputs or behave incorrectly in particular situations, such as causing a facial recognition system to misidentify a single person. A more specific type of targeted attack, known as a backdoor attack (BA) \cite{li2022backdoor}, embeds a distinct pattern or 'trigger' into portions of the training data \cite{li2022backdoor}. This trigger can be a small patch or sticker visible to humans \cite{gu2017badnets}, or a subtle perturbation in benign samples that is indistinguishable from human inspection \cite{li2020invisible}. Once the model is trained or fine-tuned on this compromised data, it behaves normally with standard inputs. However, when it encounters an input containing the covert trigger, the model exhibits malicious behavior aligned with the attacker's intentions. Note that poisoning attacks and backdoor attacks can be particularly dangerous in MU systems during the unlearning process, as they can be more stealthy and difficult to detect. A detailed review of related works and further discussion on these threats will be provided in Section \ref{sec:malicous_unlearning}.

\textbf{Membership inference attack.} Membership inference attack (MIA) allows an adversary to obtain an attack model to determine whether a specific data point was part of the model's training set \cite{shokri2017membership}. Given access to the model's predictions, attackers can exploit patterns or overfitting to infer if a particular individual's data was used in training, which can lead to privacy violations. Remarkably, MIA does not require knowledge of the target model's specific architecture or the distribution of its training data. Relying on the shadow models, a series of shadow training datasets $D'_1,\cdots,D'_k$ and disjointed shadow test datasets $T'_1,\cdots,T'_k$ can be synthesized to mimic the behavior of the target model so as to train the attack model. MIA can be used both to explore information leakage by analyzing the differences between the trained and unlearned models and to evaluate whether the target data has been successfully unlearned. This will be discussed in detail in Section \ref{sec:threats_in_unlearning} and Section \ref{sec:attacks_for_evaluation}.

\section{Workflow of Machine Unlearning Systems}
\label{sec:workflow}

In this section, we will define a unified workflow for MLaaS with unlearning services, which will aid in understanding the threats and attacks within these systems, as well as the corresponding defensive strategies.

As illustrated in Figure \ref{fig:workflow}, the structure of machine unlearning systems designed to provide unlearning services typically includes a server and various participants. In specific, the server plays key roles in the system, including:
\begin{itemize}
    \item Model developer: responsible for conducting model training based on the training data.
    \item Service provider: responsible for performing inference based on provided inputs.
\end{itemize}
Note that these roles can be allocated to different entities depending on the application scenario \cite{hu2024duty}. In addition, the roles of participants within a MU system are classified as follows:
\begin{itemize}
    \item Data contributors: responsible for providing data to construct the training dataset.
    \item Requesting users: possess the capability to submit unlearning requests.
    \item Accessible users: utilize the service for inferences.
\end{itemize}
Note that akin to the server's multi-functional role, participants within unlearning systems can embody various roles. For instance, it is common for data contributors to also have the authority to submit unlearning requests, in compliance with RTBF regulations. Additionally, the workflow of machine unlearning systems can be structured into three phases, which we categorize as follows:
\begin{itemize}
    \item Training phase: the ML model is trained using the data supplied by data contributors.
    \item Unlearning phase: the server conducts unlearning algorithms to remove the effects of certain data points from the model's knowledge base, in response to unlearning requests from requesting users.
    \item Post-unlearning phase: the server provides inference services based on unlearned models to accessible users.
\end{itemize}

In the context of machine-learning-as-a-service (MLaaS), the server may receive requests for unlearning and inference simultaneously. This indicates a possible overlap between the unlearning phase and the post-unlearning phase, where the procedures for managing different types of requests become more complex \cite{hu2023eraser}.

\section{Overview of Threats, Attacks, and Defenses in MU Systems}
\label{sec:overview}
\begin{figure*}[htbp!]
    \centering
    \includegraphics[width=0.8\linewidth]{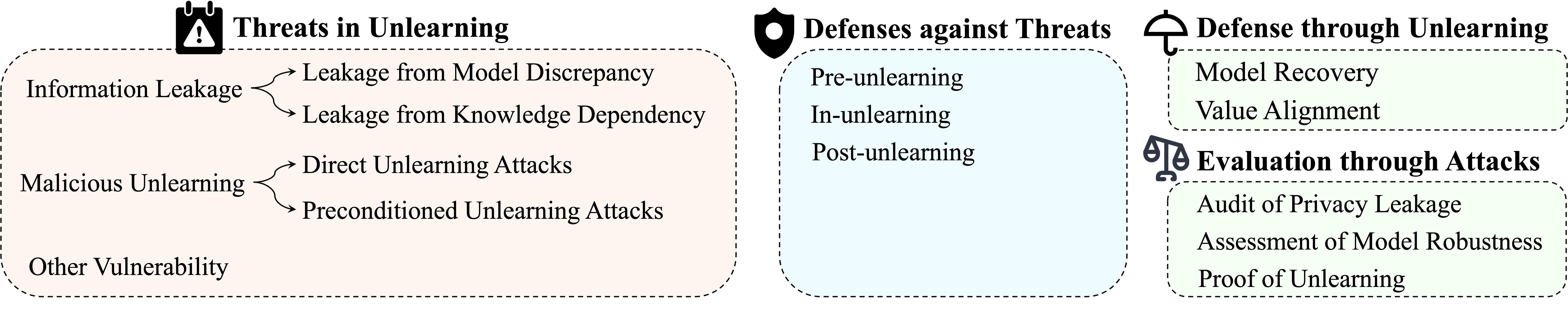}
    \caption{A taxonomy of threats, attacks, and defenses in machine unlearning.
    }
    \label{fig:taxonomy}
\end{figure*}

As mentioned earlier, vulnerabilities are present in machine unlearning systems, where adversaries can launch various attacks at different phases, posing significant threats to the safe deployment of unlearning services within MLaaS. Besides threats and attacks on machine unlearning systems, many studies show that unlearning and attacks serve dual roles. For instance, unlearning can act as a mechanism to recover models from backdoor attacks, while backdoor attacks themselves can serve as an evaluation metric for unlearning effectiveness. This reveals a complex interaction between these elements, which are essential for system safety and functionality. In this section, we offer an overview of threats, attacks, and defenses in machine unlearning systems. We explore these topics in depth in Sections \ref{sec:threats_in_unlearning}, \ref{sec:unlearning_as_defenses}, and \ref{sec:attacks_for_evaluation}, where we present a detailed taxonomy as shown in Figure \ref{fig:taxonomy}.

\noindent\textbf{Threats.} As illustrated in Figure \ref{fig:workflow}, vulnerabilities are present across all phases of the unlearning system. In the training phase, for example, data contributors can maliciously craft the training data in a way that prepares for future attacks. In the unlearning phase, requesting users have the opportunity to submit malicious unlearning requests that could be designed to achieve their goals after the unlearning process. Moreover, during the post-unlearning phase, participants with access to both the trained and the unlearned models might exploit this dual access to uncover additional information leakages, thus exposing further vulnerabilities created by the unlearning process. A more detailed discussion on the various threats that may arise is available in Section \ref{sec:threats_in_unlearning}.

\noindent\textbf{Attacks.} The threats outlined previously can be exploited to execute attacks. For instance, a data contributor might poison the training dataset, and bypass the detection by incorporating mitigation data during the training phase, which could then be removed via unlearning in the unlearning phase \cite{di2022hidden}. A requesting user could craft an unlearning request to unlearn more information than expected, thereby manipulating the system to their advantage \cite{hu2024duty}. Furthermore, an accessible user could perform enhanced membership inference attacks \cite{shokri2017membership,xie2023same,he2019model} by leveraging their access to both the trained and unlearned models, exploiting the differential insights provided by comparing the two models \cite{chen2021machine}. These attacks, along with threats to unlearning systems, are discussed in detail in Section \ref{sec:threats_in_unlearning}.

Beyond the attacks on machine unlearning systems, extensive research has demonstrated that certain prevalent attack strategies, like backdoor attacks \cite{liu2018trojaning} and membership inference attacks \cite{shokri2017membership}, can be adapted for unlearning evaluations, such as privacy leakage audit \cite{chen2021machine}, model robustness assessment \cite{hu2024duty}, and proof of unlearning \cite{sommer2020towards}. The analysis of attacks used for evaluating unlearning is provided in Section \ref{sec:unlearning_as_defenses}.

\noindent\textbf{Defenses.} Potential defense methods against attacks on MU systems are explored in conjunction with attack strategies. For example, performing membership checks on data targeted for unlearning can ensure that the data specified in unlearning requests genuinely exists in the training dataset \cite{hu2024duty}. Implementing differential privacy \cite{dwork2006differential,zhang2024bounded} can help prevent membership inference attacks, with a sacrifice of model performance \cite{chen2021machine}. Moreover, anomaly detection through model monitoring in the unlearning phase can effectively identify potential issues \cite{zhang2023exploiting}. A corresponding discussion, along with an analysis of threats and attacks, is given in Section \ref{sec:threats_in_unlearning}.

Additionally, unlearning can serve as a method of defense. It is capable of removing backdoors from trained machine learning models \cite{liu2022backdoor,cao2023fedrecover,wang2019neural}, and can be employed to achieve value alignment, thereby providing a defense mechanism against potential AI safety issues \cite{lu2022quark,liu2024rethinking,liu2024model}. The dual role of unlearning as a defense mechanism is analyzed in Section \ref{sec:unlearning_as_defenses}.

In summary, threats, attacks, and defenses within machine unlearning systems exhibit a complex relationship, revealing an intricate interplay among these elements in upholding system functionality and safety.

\section{Threat Models}
\label{sec:threat_models}

In this section, we define the threat models based on our proposed unified MU workflow described in Figure \ref{fig:workflow}, regarding the role of attackers in unlearning services, the attackers' goals, their knowledge of the MU system, and the phases during which attacks occur. This analysis of threat models will provide a deeper understanding of the existing related works discussed in the subsequent sections.

\subsection{Attack Roles}
Threats and attacks within the machine unlearning workflow can arise from participants adopting various adversarial roles. These roles can be categorized into three distinct classes, as presented in Section \ref{sec:introduction}, outlined below:

\begin{itemize}
    \item \textbf{Data contributors (R1)}: Individuals responsible for providing data to construct the training dataset.
    \item \textbf{Requesting users (R2)}: Individuals who have the ability to submit unlearning requests.
    \item \textbf{Accessible users (R3)}: Individuals with access to the model service, enabling them to utilize the model for making inferences.
\end{itemize}

Note that attackers within unlearning systems can fulfill various attack roles. A requesting user, for example, usually has access to the model service for inference purposes. Similarly, requesting users can also serve as data contributors. Generally, the more roles an attacker holds, the stronger their potential for launching effective attacks becomes.

\subsection{Attack Goals}
Attack goals within unlearning systems also vary and can generally be categorized as follows:

\begin{itemize}
    \item \textbf{Untargeted (G1)}: Intend to mislead the unlearned model into generating incorrect predictions without targeting a specific outcome.
    \item \textbf{Targeted (G2)}: Aim at inducing the unlearned model to produce incorrect predictions or behave in a specific, predetermined way.
    \item \textbf{Privacy leakage (G3)}: Seeks to extract additional information about the data requested to be unlearned.
    \item \textbf{Others (G4)}: Goals may include increasing the unlearning process's computational cost, impacting the unlearned model's fairness and utility, etc.
\end{itemize}

\subsection{Adversarial Knowledge}
Additionally, attackers may possess varying levels of adversarial knowledge for an attack, which can be classified into three categories. Generally, the more knowledge an attacker has, the more potent the attack becomes.

\begin{itemize}
    \item \textbf{White-box (K1)}: The attacker possesses comprehensive knowledge of the model's architecture, parameters, training and unlearning algorithms, and data.
    \item \textbf{Grey-box (K2)}: The attacker has access to some elements such as the model's architecture, parameters, training and unlearning algorithms, or data, but not to all of them.
    \item \textbf{Black-box (K3)}: The attacker has no knowledge of the model, including its architecture and parameters, and cannot access or modify the model's training process.
\end{itemize}

\subsection{Attack Phases}
Attacks and threats can occur at different stages within an ML system with unlearning services, and can be categorized into the following three phases:

\begin{itemize}
    \item \textbf{Training phase (P1)}: Attackers, serving as data contributors, may manipulate or craft the training dataset to prepare for an attack during the later phases.
    \item \textbf{Unlearning phase (P2)}:  Attackers may submit malicious unlearning requests to initiate attacks, which may or may not be based on preparations in the training phase.
    \item \textbf{Post-unlearning phase (P3)}: Attackers may launch attacks leveraging the unlearned model and information acquired during the previous phases.
\end{itemize}




\section{Threats in Unlearning}
\label{sec:threats_in_unlearning}

In this section, we explore the threats inherent to machine unlearning systems. Our focus encompasses the threats of information leakage through machine unlearning, the impact of malicious unlearning practices, and other vulnerabilities that may arise within unlearning systems. We offer a thorough analysis of their methodologies based on their respective threat models (see Section \ref{sec:threat_models} for more details) and discuss potential defense mechanisms to mitigate these risks.

\subsection{Information Leakage}
\label{sec:information_leakage}

As previously noted, machine unlearning can introduce specific forms of information leakage. This leakage often stems from (i) discrepancies between the trained and unlearned models, or from (ii) the insights derived from the dependency between the model and external knowledge. In this section, we will explore these two sources of information leakage. It is important to emphasize that our discussion is centered on information leakage uniquely associated with machine unlearning, rather than the general leakage issues found in standard ML systems.

\noindent\textbf{Leakage from model discrepancy.} Model discrepancy refers to the differences observed between the trained model and the unlearned model. Attackers can take advantage of this variance to extract additional information about the unlearned data. This could involve determining the membership of specific unlearned data points \cite{chen2021machine, lu2022label, hu2023eraser, gao2022deletion} or attempting to reconstruct the data that was intended to be unlearned \cite{gao2022deletion}.

In particular, \cite{chen2021machine} utilizes a membership inference attack strategy to determine whether a given input was part of the unlearned data. For a specific input, both the trained and unlearned models produce confidence values represented as vectors. By strategically combining these vectors of confidence values, an independent attack model can be trained to deduce the presence of the input within the unlearned data, essentially inferring its membership. Compared to \cite{chen2021machine}, the membership inference attack outlined in \cite{lu2022label} necessitates less information, relying solely on the Top-1 value of the confidence vectors, namely, the inferred label. To accomplish this, the attacker strategically perturbs the input to observe the changes in the outputs of both the trained and unlearned models, thereby deducing the membership of the input. In \cite{gao2022deletion}, a general framework for inference and reconstruction attacks is formally established, discussing various instances across different machine learning tasks. Data construction attacks are also investigated in \cite{hu2024learn}, where the model discrepancy is used as estimated gradients. Based on this estimation, attacks through deep leakage from gradients \cite{geiping2020inverting, zhu2019deep} can be employed to achieve data reconstruction. Additionally, \cite{hu2024learn} demonstrates that label inference attacks can be effective even with only access to prediction services, without needing the model parameters, in a black-box setting. The information leakage from model discrepancy has been shown to be effective for textual tasks in \cite{du2024textual}. Based on the proposed attack methods, attackers can infer the membership of unlearned data in a black-box setting and reconstruct the unlearned data with white-box access to the two versions of the model.

Note that the privacy leakages mentioned above occur in the post-unlearning phase when accessible users can interact with the model service for inferences. However, \cite{hu2023eraser} highlights a new privacy risk from the Rignt-To-Be-Forgetten perspective in an MLaaS setting where the unlearning phase for one request and the post-unlearning phase for another inference request may overlap, i.e., inference requests and unlearning requests may arrive at the server within a very close timeframe. In such scenarios, prioritizing the ``unlearning-request-first" could lead to service obsolescence, whereas handling the ``inference-request-first" enables an attacker to gain knowledge about a specific input, which can be exploited by attacks like membership inference attacks. To counter this threat, strategies have been developed to evaluate the urgency of processing unlearning requests immediately. This evaluation is based on whether the unlearning disrupts the consistency between the trained and unlearned models.

\noindent\textbf{Leakage from knowledge dependency.} Knowledge dependency stems from the natural relationships between the model and external knowledge sources, such as the association between successive unlearning requests and the unlearned model \cite{chourasia2023forget}, as well as the correlation between open-source model parameters and the embedding of prompts in LLMs \cite{schwinn2024soft}. Leveraging these dependencies to gain extra knowledge, attackers may initiate various attacks leading to information leakage.

In \cite{chourasia2023forget}, two types of knowledge dependency are described within machine unlearning systems. 
The first one involves adaptive requests \cite{gupta2021adaptive}, indicating that an unlearning request to remove certain data points may influence the removal of other data points. This relationship can be exploited through a series of successive unlearning requests, gathering sufficient knowledge to ascertain the membership of specific data. The second instance of knowledge dependency arises from unlearning algorithms caching partial computations to expedite processing. This method can unintentionally reveal information about data meant to be deleted across multiple releases. The dependency between the parameters of unlearned large models and the embedding of prompts is investigated in \cite{schwinn2024soft}. By introducing adversarial perturbations to the embeddings of prompt tokens, it is possible to instruct the large model to reveal knowledge that was supposed to be erased through unlearning.

\noindent\textbf{Defenses.} 
To mitigate privacy attacks and subsequent information leakage, the essential strategy is to limit the exploitable knowledge for attackers. For model discrepancy-related leakage, reducing the amount of information provided by the models, such as only showing top-k or top-1 confidence scores \cite{chen2021machine}, applying techniques like label smoothing and differential privacy \cite{chen2021machine,lu2022label,gao2022deletion}, or strategic postponed unlearning \cite{hu2023eraser}, are effective approaches. Regarding leakage from knowledge dependency, solutions vary. Specifically, prevent unlearning algorithms from caching can address leaks from cached data, and redefining privacy metrics for adaptive requests can help measure and control privacy risks under those scenarios.

\begin{figure*}[t]
\centering
    \begin{subfigure}{0.45\textwidth}
    \centering
    \includegraphics[width=0.9\linewidth]{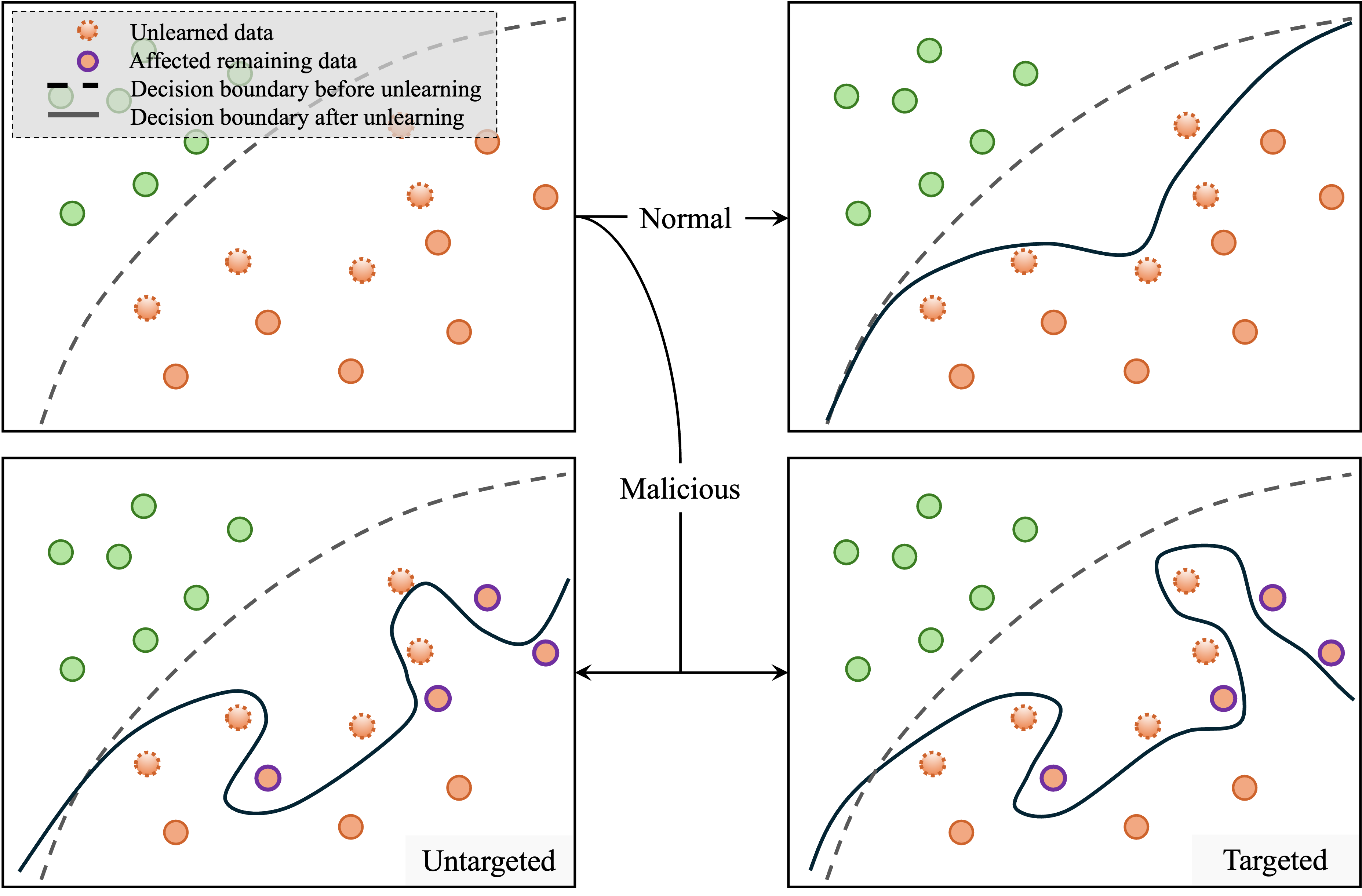}
    \caption{Direct unlearning attacks.}
    \label{fig:direct}
    \end{subfigure}
    \begin{subfigure}{0.45\textwidth}
    \centering
    \includegraphics[width=0.9\linewidth]{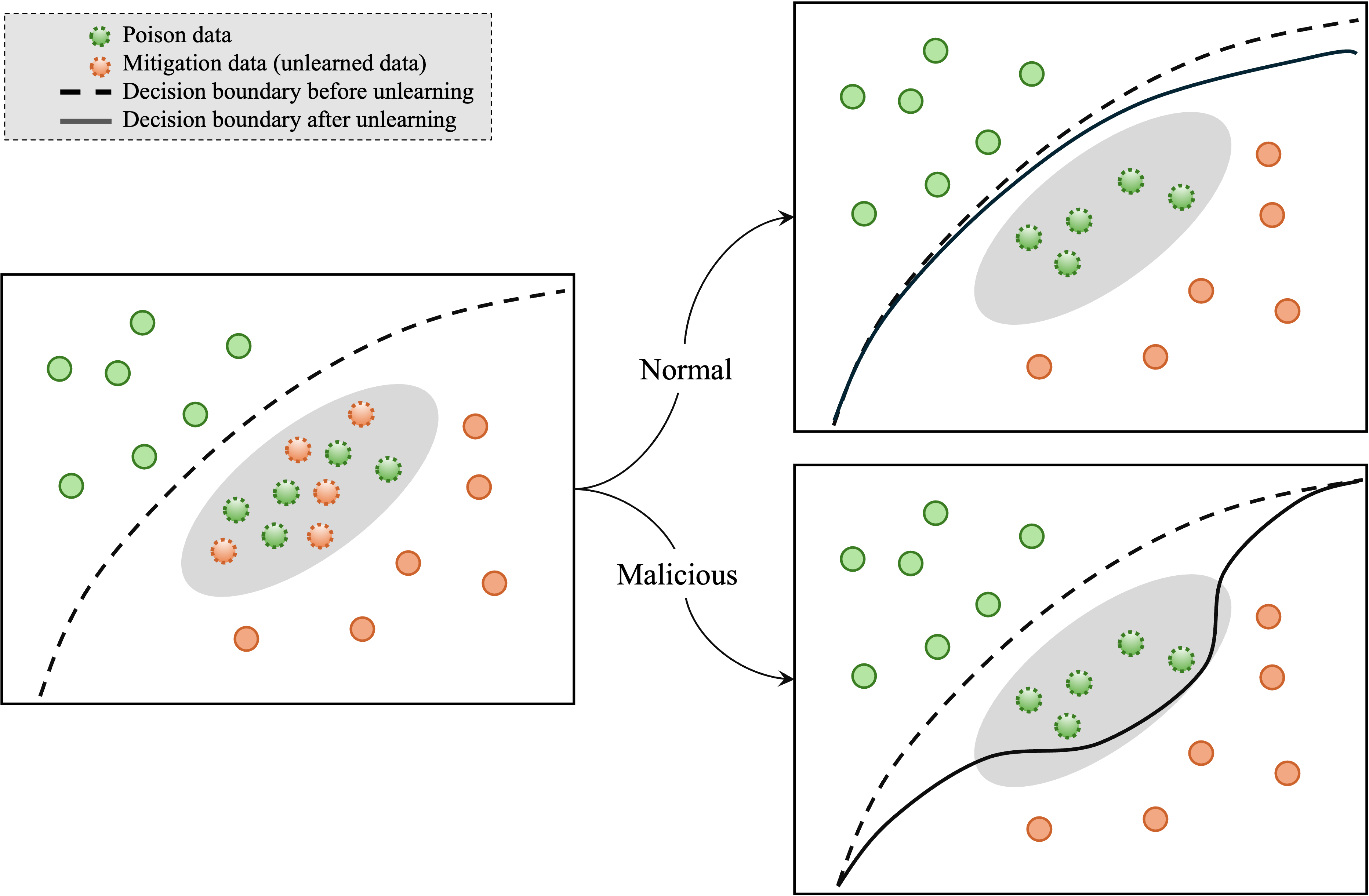}
    \caption{Preconditioned unlearning attacks.}
    \label{fig:preconditioned}
    \end{subfigure}
\caption{
An illustration of direct and preconditioned unlearning attacks via decision boundary manipulation. (a) \textit{Direct unlearning attacks} occur solely in the unlearning phase, with no manipulation of training data. (b) \textit{Preconditioned unlearning attacks} involve strategic manipulation of training data.}
\label{fig:decision_boundary}
\end{figure*}

\subsection{Malicious Unlearning}
\label{sec:malicous_unlearning}

As noted earlier, attackers can engage in malicious unlearning by submitting crafted unlearning requests during the unlearning phase. It is important to note that such an attack happens in the unlearning phase, but the attacker may or may not need some preparation, e.g., manipulating the training data, in the training phase. We call malicious unlearning attacks that do not need preparation ``direct unlearning attacks" and those that require preparation ``preconditioned unlearning attacks". Next, we will explore the methods and principles of these two types of attacks.

\noindent\textbf{Direct unlearning attacks.} Direct unlearning attacks occur solely in the unlearning phase, without requiring the manipulation of training data during the training phase, while attackers may or may not possess knowledge of the training data. These attacks are designed to be either untargeted, aiming to degrade the overall performance of the model after unlearning \cite{hu2024duty}, or targeted, with the intention of causing the unlearned model to misclassify target inputs with predefined features \cite{zhao2024static}.

Specifically, \cite{hu2024duty} demonstrates how untargeted attacks can be conducted through over-unlearning techniques in a black-box setting. This is based on the critical observation that models struggle to predict data points close to the decision boundary, as even minor changes in the input could result in different predictions. As depicted in Figure \ref{fig:direct}, after a malicious unlearning process, a number of data points end up closer to the decision boundary than in normal unlearning scenarios. Consequently, these data points are more prone to misclassification compared to the normal case. Hence, the unlearned model obtained after malicious unlearning exhibits lower classification accuracy than a model that has undergone normal unlearning. For crafting such malicious unlearning requests, attackers leverage the model's inference service to acquire knowledge about the trained model. This knowledge enables them to formulate malicious unlearning requests employing methods of adversarial perturbation \cite{carlini2017towards,chen2017zoo}.

Distinct from the untargeted attacks described in \cite{hu2024duty}, targeted attacks can be executed with additional knowledge of the training data in a grey-box or white-box setting, as detailed in \cite{zhao2024static}. Unlike the approach in \cite{hu2024duty} that shifts the decision boundary closer to remaining data points, targeted attacks manipulate the decision boundary to selectively cross through remaining data points, classifying specific data points with predefined features into a target class. As outlined in \cite{zhao2024static}, to generate these malicious unlearning requests, attackers exploit the model's inference service along with training data to which they have access. They address an optimization problem to identify the optimal perturbation and the data points to be perturbed. These selected data points, once perturbed, are then submitted as part of the malicious unlearning requests.

\noindent\textbf{Preconditioned unlearning attacks.} Unlike direct unlearning attacks, preconditioned unlearning attacks take a more strategic approach by manipulating training data during the training phase. These attacks are typically designed to carry out complex and stealthy targeted attacks.

As depicted in Figure \ref{fig:preconditioned}, a typical preconditioned unlearning attack is executed through the following steps:

\begin{enumerate}
    \item The attackers incorporate poisoned data $D_p$ and mitigation data $D_m$ into a clean dataset $D_c$, to ensemble a training dataset $D=D_p \cup D_m \cup D_c$.
    \item The server trains the ML model based on $D$ and returns the model $\mathcal{M}$.
    \item The attackers submit requests to unlearn the mitigation data $D_m$.
    \item Upon receiving the requests, the server carries out the unlearning process and returns the unlearned model $\hat{\mathcal{M}}$, which effectively corresponds to a model trained on the dataset $D_p \cup D_c$.
\end{enumerate}

Thus, after the unlearning process removes the mitigation data, the model is left vulnerable to the poisoned data inserted during the training phase. This exposes the unlearned model to the influence of the initial poisoning, culminating in a successful preconditioned unlearning attack.

A number of studies have successfully executed preconditioned unlearning attacks, including \cite{zhang2023exploiting,di2022hidden,qian2023towards}. The primary distinction between \cite{di2022hidden} and \cite{zhang2023exploiting} lies in their approaches to generating poisoned and mitigation data during the training phase. In \cite{di2022hidden}, poison generation is approached as a bilevel optimization problem, as described in \cite{geiping2020witches}, with mitigation data produced either through label flipping or gradient matching to neutralize the poison's effect. On the other hand, \cite{zhang2023exploiting} outlines two methodologies for crafting the training dataset: (i) generating poisoned data by perturbing sampled clean data, with mitigation data created using the same sampled perturbed clean data but with ground-truth labels, akin to the label flipping technique in \cite{geiping2020witches}, and (ii) a different strategy for poisoned data generation inspired by Badnets \cite{gu2019badnets}, where the mitigation data contains a trigger that exerts a stronger influence on the model, hence concealing the effects of the poisoned data. Camouflaged sample generation based on data influence is explored in \cite{huang2024uba}, along with a comprehensive performance evaluation of various machine unlearning methods.

In addition to the typical preconditioned unlearning attacks detailed in \cite{zhang2023exploiting,di2022hidden}, another strategy is introduced in \cite{qian2023towards} where attackers deliberately perturb certain pieces of training data, which are subsequently used in model training. In the unlearning phase, attackers strategically submit requests for the removal of the perturbed data points. This approach takes advantage of the knowledge gained from the model's inference service and the perturbed training data, thereby executing an effective form of preconditioned unlearning attack. Another type of preconditioned unlearning attack is proposed in \cite{ma2024releasing}, where attackers deliberately generate training data. The overall model usability will degrade after unlearning these data.

\noindent\textbf{Defenses.} Defensive strategies to counteract malicious unlearning have been explored in the studies previously mentioned. A notable observation is a significant distinction in defense methods when addressing direct versus preconditioned unlearning attacks. The distinction arises because, in direct unlearning attacks, the data submitted for unlearning are not part of the original training dataset. In contrast, in preconditioned unlearning attacks, the data requested to be unlearned are indeed from the training dataset. For instance, to prevent malicious unlearning requests in direct unlearning attacks, one can compare the Hash value of the unlearned data and training data or make the membership inference as pointed out in \cite{hu2024duty,ma2024releasing}. 
In the case of preventing malicious unlearning requests during preconditioned unlearning attacks, membership checking is ineffective because the data being unlearned is part of the original training dataset. Consequently, \cite{zhang2023exploiting} suggests detection strategies based on metrics that compare the model's performance on clean data versus mitigation data. If the discrepancy exceeds a certain threshold, the unlearning request may be denied. On a different note, \cite{qian2023towards} proposes not directly rejecting unlearning requests but instead dropping malicious unlearned data from requests that exhibit a gradient dissimilarity compared to other instances within their class.

\subsection{Other Vulnerabilities}

In addition to privacy leakages resulting from unlearning and threats induced by malicious unlearning practices, various other vulnerabilities within machine unlearning have been explored from multiple perspectives.

In \cite{marchant2022hard}, the concept of a ``slow-down attack" is explored, identifying it as a type of poisoning attack aimed at decelerating the unlearning process. This attack is particularly relevant in the context of certified removal \cite{guo2019certified}, a system that decides between conducting a full retraining or an approximate update based on specific triggers. By strategically crafting poisoned data in the training data, one can minimize the interval or the number of unlearning requests processed through approximate updates before necessitating full retraining. Since a full retrain is considerably more time-consuming than an approximate update, such a ``slow-down attack" significantly diminishes the overall efficiency of unlearning. The target of creating this poisoned data is formulated as a bilevel optimization problem \cite{mei2015using}, where locally optimal solutions can be obtained using gradient-based methodologies.

Other studies highlight that unlearning algorithms themselves might introduce side effects in specific domain applications. For example, \cite{kadhe2023fairsisa} discusses the impact on fairness in LLMs, noting that ensemble unlearning methods, like those detailed in \cite{bourtoule2021machine}, can affect fairness. A proposed mitigation approach involves adjusting the LLMs' outputs using bias mitigation techniques outlined in \cite{soares2022your}. Meanwhile, \cite{tan2024unlink}, focusing on Graph Neural Networks (GNNs), reveals that unlearning algorithms, such as \cite{cheng2023gnndelete}, can lead to over-forgetting. This phenomenon occurs when the unlearning process inadvertently removes more information than necessary, significantly reducing the performance of the remaining edges. Since over-forgetting stems from the design of the unlearning algorithm itself, \cite{tan2024unlink} suggests modifying the unlearning algorithms to prevent such outcomes.

\newcolumntype{M}[1]{>{\vspace{2pt}}m{#1}<{\vspace{2pt}}}

\begin{table*}[htbp!]
\renewcommand{\arraystretch}{1.2}
\centering
\resizebox{\textwidth}{!}{
\begin{tabular}{ll|c|p{0.15cm}m{0.15cm}p{0.17cm}|m{0.15cm}m{0.15cm}m{0.15cm}p{0.17cm}|m{0.15cm}m{0.15cm}p{0.17cm}|m{0.15cm}m{0.15cm}p{0.16cm}|M{10.5em}|M{12.5em} }

\toprule 

\multicolumn{2}{c|}{} & 
    & \multicolumn{3}{c|}{\textbf{Roles}} 
    & \multicolumn{4}{c|}{\textbf{Goals}} 
    & \multicolumn{3}{c|}{\textbf{Knowledge}} 
    & \multicolumn{3}{c|}{\textbf{Phases}} 
    & \multicolumn{1}{c|}{} 
    & \multicolumn{1}{c}{} \\
\multicolumn{2}{c|}{\multirow{-2}{*}{\textbf{Threats}}} 
    & \multirow{-2}{*}{\textbf{Ref.}} 
    & R1 & R2 & R3 
    & G1 & G2 & G3 & G4 
    & K1 & K2 & K3                        
    & P1 & P2 & P3                        
    & \multicolumn{1}{c|}{\multirow{-2}{*}{\textbf{Techniques}}}
    & \multicolumn{1}{c}{\multirow{-2}{*}{\textbf{Defenses}}} \\ 
\midrule 

    \multicolumn{1}{l|}{} &        
        &\cellcolor[HTML]{EFEFEF} \cite{chen2021machine}
        &\cellcolor[HTML]{EFEFEF} &\cellcolor[HTML]{EFEFEF}\checkmark  &\cellcolor[HTML]{EFEFEF}\checkmark
        &\cellcolor[HTML]{EFEFEF} &\cellcolor[HTML]{EFEFEF}  &\cellcolor[HTML]{EFEFEF}\checkmark &\cellcolor[HTML]{EFEFEF}
        &\cellcolor[HTML]{EFEFEF} &\cellcolor[HTML]{EFEFEF}  &\cellcolor[HTML]{EFEFEF} \checkmark
        &\cellcolor[HTML]{EFEFEF} &\cellcolor[HTML]{EFEFEF}  &\cellcolor[HTML]{EFEFEF} \checkmark
        &\cellcolor[HTML]{EFEFEF} \smallsix{Membership inference attacks}
        &\cellcolor[HTML]{EFEFEF} \smallsix{\makecell[l]{Limiting output knowledge \tiny{(S3)}; \\[1ex] Label smoothing \tiny{(S3)}; \\[1ex] Differential privacy \tiny{(S3)}}} \\ 
    \multicolumn{1}{l|}{} &
        &\cite{lu2022label}
        & &\checkmark & \checkmark
        & & & \checkmark & 
        & & & \checkmark
        & & & \checkmark
        & \smallsix{Membership inference attacks}
        & \smallsix{\makecell[l]{Label smoothing \tiny{(S3)}; \\[1ex] Differential privacy \tiny{(S3)}}} \\
    \multicolumn{1}{l|}{} &
        &\cellcolor[HTML]{EFEFEF} \cite{gao2022deletion}
        &\cellcolor[HTML]{EFEFEF} &\cellcolor[HTML]{EFEFEF} &\cellcolor[HTML]{EFEFEF} \checkmark
        &\cellcolor[HTML]{EFEFEF} &\cellcolor[HTML]{EFEFEF} &\cellcolor[HTML]{EFEFEF} \checkmark &\cellcolor[HTML]{EFEFEF} 
        &\cellcolor[HTML]{EFEFEF} &\cellcolor[HTML]{EFEFEF} &\cellcolor[HTML]{EFEFEF} \checkmark
        &\cellcolor[HTML]{EFEFEF} &\cellcolor[HTML]{EFEFEF} &\cellcolor[HTML]{EFEFEF} \checkmark
        &\cellcolor[HTML]{EFEFEF} \smallsix{\makecell[l]{Deletion inference attacks; \\[1ex] Deletion reconstruction attacks}}
        &\cellcolor[HTML]{EFEFEF} \smallsix{Differential privacy \tiny{(S3)}} \\
    \multicolumn{1}{l|}{} & 
        & \cite{hu2023eraser}
        & & & \checkmark
        & & & \checkmark & 
        & & &\checkmark
        & & \checkmark&\checkmark
        & \smallsix{Obsolescent inference services}
        & \smallsix{Strategic postponed unlearning \tiny{(S2,S3)}}  \\
    \multicolumn{1}{l|}{} &
        &\cellcolor[HTML]{EFEFEF} \cite{hu2024learn}
        &\cellcolor[HTML]{EFEFEF} &\cellcolor[HTML]{EFEFEF} &\cellcolor[HTML]{EFEFEF} \checkmark
        &\cellcolor[HTML]{EFEFEF} &\cellcolor[HTML]{EFEFEF}  &\cellcolor[HTML]{EFEFEF} \checkmark &\cellcolor[HTML]{EFEFEF} 
        &\cellcolor[HTML]{EFEFEF} \checkmark &\cellcolor[HTML]{EFEFEF} &\cellcolor[HTML]{EFEFEF} \checkmark
        &\cellcolor[HTML]{EFEFEF} &\cellcolor[HTML]{EFEFEF} &\cellcolor[HTML]{EFEFEF} \checkmark
        &\cellcolor[HTML]{EFEFEF} \smallsix{\makecell[l]{Unlearning inversion attacks}}
        &\cellcolor[HTML]{EFEFEF} \smallsix{\makecell[l]{Parameter obfuscation \tiny{(S1)}; \\[1ex] Model pruning \tiny{(S3)}; \\[1ex] Fine-tuning \tiny{(S3)}}} \\
    \multicolumn{1}{l|}{} & 
        & \cite{bertran2024reconstruction}
        & & & \checkmark
        & & & \checkmark & 
        &\checkmark & &
        & & &\checkmark
        & \smallsix{Reconstruction attacks}
        & \smallsix{Differential privacy \tiny{(S3)}}  \\
    \multicolumn{1}{l|}{} & \multirow{-11}{*}{\begin{tabular}[c]{@{}l@{}}Model\\ Discrepancy\end{tabular}} 
        &\cellcolor[HTML]{EFEFEF} \cite{du2024textual}
        &\cellcolor[HTML]{EFEFEF} &\cellcolor[HTML]{EFEFEF} \checkmark &\cellcolor[HTML]{EFEFEF} \checkmark
        &\cellcolor[HTML]{EFEFEF} &\cellcolor[HTML]{EFEFEF}            &\cellcolor[HTML]{EFEFEF} \checkmark  &\cellcolor[HTML]{EFEFEF} 
        &\cellcolor[HTML]{EFEFEF} \checkmark &\cellcolor[HTML]{EFEFEF} &\cellcolor[HTML]{EFEFEF} \checkmark
        &\cellcolor[HTML]{EFEFEF} &\cellcolor[HTML]{EFEFEF} &\cellcolor[HTML]{EFEFEF} \checkmark
        &\cellcolor[HTML]{EFEFEF} \smallsix{Textual unlearning leakage attack}
        &\cellcolor[HTML]{EFEFEF} \smallsix{NA}\\
    
\cline{2-18} \\[-1.1em]
    
    \multicolumn{1}{l|}{} &
        & \cite{chourasia2023forget}
        & & \checkmark  & \checkmark    
        & & &\checkmark &   
        & &\checkmark &    
        & & \checkmark &\checkmark    
        & \smallsix{\makecell[l]{Submitting adaptive requests; \\[1ex] Storing secret models}}   
        & \smallsix{\makecell[l]{Regulated unlearning algorithms \tiny{(S1)}; \\[1ex] Redefined certified removal \tiny{(S1)}}}  \\
\multicolumn{1}{l|}{\multirow{-15}{*}{\begin{tabular}[c]{@{}l@{}}Information\\ Leakage\end{tabular}}}    
    & \multirow{-3}{*}{\begin{tabular}[c]{@{}l@{}}Knowledge\\ Dependency\end{tabular}}                   
        &\cellcolor[HTML]{EFEFEF} \cite{schwinn2024soft}
        &\cellcolor[HTML]{EFEFEF} &\cellcolor[HTML]{EFEFEF} &\cellcolor[HTML]{EFEFEF}\checkmark    
        &\cellcolor[HTML]{EFEFEF} &\cellcolor[HTML]{EFEFEF} &\cellcolor[HTML]{EFEFEF}\checkmark &\cellcolor[HTML]{EFEFEF}  \checkmark
        &\cellcolor[HTML]{EFEFEF}\checkmark &\cellcolor[HTML]{EFEFEF} &\cellcolor[HTML]{EFEFEF}    
        &\cellcolor[HTML]{EFEFEF} &\cellcolor[HTML]{EFEFEF} &\cellcolor[HTML]{EFEFEF}\checkmark
        &\cellcolor[HTML]{EFEFEF} \smallsix{Adversarial embeddings}
        &\cellcolor[HTML]{EFEFEF} \smallsix{NA}  \\ 
\midrule  
    
    \multicolumn{1}{l|}{} &
        & \cite{hu2024duty}
        & & \checkmark  & \checkmark  
        &\checkmark  &  &  & 
        & & & \checkmark
        & & \checkmark  &
        & \smallsix{Over-unlearning}
        & \smallsix{\makecell[l]{Membership checking \tiny{(S1)}; \\[1ex] Model monitoring \tiny{(S2)}}} \\
    \multicolumn{1}{l|}{} & 
        &\cellcolor[HTML]{EFEFEF}  \cite{zhao2024static}
        &\cellcolor[HTML]{EFEFEF}  &\cellcolor[HTML]{EFEFEF}  \checkmark &\cellcolor[HTML]{EFEFEF}  \checkmark    
        &\cellcolor[HTML]{EFEFEF}  \checkmark &\cellcolor[HTML]{EFEFEF}  \checkmark &\cellcolor[HTML]{EFEFEF}  &\cellcolor[HTML]{EFEFEF}    
        &\cellcolor[HTML]{EFEFEF} \checkmark &\cellcolor[HTML]{EFEFEF}  \checkmark &\cellcolor[HTML]{EFEFEF}  \checkmark   
        &\cellcolor[HTML]{EFEFEF}  &\cellcolor[HTML]{EFEFEF}  \checkmark &\cellcolor[HTML]{EFEFEF}     
        &\cellcolor[HTML]{EFEFEF}  \smallsix{Perturbing unlearning requests}
        &\cellcolor[HTML]{EFEFEF}  \smallsix{NA}  \\
    \multicolumn{1}{l|}{} & \multirow{-4}{*}{\begin{tabular}[c]{@{}l@{}}Direct \\ Unlearning\\ Attacks\end{tabular}}    
        &  \cite{liu2024backdoor}
        & \checkmark  & \checkmark  & \checkmark    
        &             & \checkmark &               &    
        & \checkmark  &             &  \checkmark   
        & \checkmark  & \checkmark  &     
        &  \smallsix{Backdoor attacks}
        &  \smallsix{NA}  \\
\cline{2-18} \\[-1.1em]
    \multicolumn{1}{l|}{} & 
        &\cellcolor[HTML]{EFEFEF}   \cite{liu2024backdoor}
        &\cellcolor[HTML]{EFEFEF}  \checkmark  &\cellcolor[HTML]{EFEFEF}  \checkmark  &\cellcolor[HTML]{EFEFEF}  \checkmark    
        &\cellcolor[HTML]{EFEFEF}              &\cellcolor[HTML]{EFEFEF}  \checkmark &\cellcolor[HTML]{EFEFEF}                &\cellcolor[HTML]{EFEFEF}     
        &\cellcolor[HTML]{EFEFEF}  \checkmark  &\cellcolor[HTML]{EFEFEF}              &\cellcolor[HTML]{EFEFEF}   \checkmark   
        &\cellcolor[HTML]{EFEFEF}  \checkmark  &\cellcolor[HTML]{EFEFEF}  \checkmark  &\cellcolor[HTML]{EFEFEF}      
        &\cellcolor[HTML]{EFEFEF}   \smallsix{Backdoor attacks}
        &\cellcolor[HTML]{EFEFEF}   \smallsix{NA}  \\
    \multicolumn{1}{l|}{} &
        &\cite{di2022hidden}
        & \checkmark  &\checkmark  &   
        &             & \checkmark &   &  
        &             & \checkmark &   
        & \checkmark  & \checkmark &   
        &\smallsix{Camouﬂaged attacks}   
        &\smallsix{NA}  \\
    \multicolumn{1}{l|}{} &
        &\cellcolor[HTML]{EFEFEF} \cite{zhang2023exploiting}
        &\cellcolor[HTML]{EFEFEF} \checkmark &\cellcolor[HTML]{EFEFEF} \checkmark &\cellcolor[HTML]{EFEFEF}    
        &\cellcolor[HTML]{EFEFEF} &\cellcolor[HTML]{EFEFEF} \checkmark &\cellcolor[HTML]{EFEFEF} &\cellcolor[HTML]{EFEFEF}   
        &\cellcolor[HTML]{EFEFEF} &\cellcolor[HTML]{EFEFEF} &\cellcolor[HTML]{EFEFEF} \checkmark   
        &\cellcolor[HTML]{EFEFEF} \checkmark &\cellcolor[HTML]{EFEFEF} \checkmark &\cellcolor[HTML]{EFEFEF}     
        &\cellcolor[HTML]{EFEFEF} \smallsix{Camouﬂaged attacks}   
        &\cellcolor[HTML]{EFEFEF} \smallsix{Model monitoring \tiny{(S2)}}  \\
    \multicolumn{1}{l|}{} &
        & \cite{qian2023towards}
        & & \checkmark & \checkmark    
        & \checkmark & \checkmark & &
        &\checkmark & & \checkmark
        & \checkmark & \checkmark &
        & \smallsix{Perturbing training data}    
        & \smallsix{Malicious requests detection \tiny{(S1)}}   \\  
\multicolumn{1}{l|}{\multirow{-11}{*}{\begin{tabular}[c]{@{}l@{}}Malicious\\ Unlearning\end{tabular}}} 
    & \multirow{-6}{*}{\begin{tabular}[c]{@{}l@{}}Preconditioned\\ Unlearning\\ Attacks\end{tabular}} 
        &\cellcolor[HTML]{EFEFEF} \cite{ma2024releasing}
        &\cellcolor[HTML]{EFEFEF} \checkmark &\cellcolor[HTML]{EFEFEF} \checkmark &\cellcolor[HTML]{EFEFEF}    
        &\cellcolor[HTML]{EFEFEF} \checkmark &\cellcolor[HTML]{EFEFEF}            &\cellcolor[HTML]{EFEFEF} &\cellcolor[HTML]{EFEFEF}   
        &\cellcolor[HTML]{EFEFEF}            &\cellcolor[HTML]{EFEFEF}            &\cellcolor[HTML]{EFEFEF} \checkmark   
        &\cellcolor[HTML]{EFEFEF} \checkmark &\cellcolor[HTML]{EFEFEF} \checkmark &\cellcolor[HTML]{EFEFEF}     
        &\cellcolor[HTML]{EFEFEF} \smallsix{Unlearning usability attack}   
        &\cellcolor[HTML]{EFEFEF} \smallsix{Model monitoring \tiny{(S2)}}  \\
    \multicolumn{1}{l|}{} &
        & \cite{huang2024uba}
        & \checkmark & \checkmark &     
        & & \checkmark & &
        & & \checkmark & \checkmark
        & \checkmark & \checkmark &
        & \smallsix{Camouﬂaged attacks}    
        & \smallsix{NA} \\
\midrule  

    \multicolumn{2}{l|}{}   
        & \cite{marchant2022hard}  
        &\checkmark &\checkmark &    
        & & & & \checkmark
        &\checkmark &\checkmark &    
        &\checkmark & \checkmark& 
        &  \smallsix{Poisoning training data}
        &  \smallsix{NA} \\
    \multicolumn{2}{l|}{}
        &\cellcolor[HTML]{EFEFEF}  \cite{kadhe2023fairsisa}
        &\cellcolor[HTML]{EFEFEF}  \checkmark &\cellcolor[HTML]{EFEFEF}  &\cellcolor[HTML]{EFEFEF}     
        &\cellcolor[HTML]{EFEFEF}  &\cellcolor[HTML]{EFEFEF}  &\cellcolor[HTML]{EFEFEF}  &\cellcolor[HTML]{EFEFEF}  \checkmark  
        &\cellcolor[HTML]{EFEFEF}  \checkmark &\cellcolor[HTML]{EFEFEF}  &\cellcolor[HTML]{EFEFEF}     
        &\cellcolor[HTML]{EFEFEF}  &\cellcolor[HTML]{EFEFEF} \checkmark &\cellcolor[HTML]{EFEFEF}     
        &\cellcolor[HTML]{EFEFEF}   \smallsix{Ensemble unlearning}  
        &\cellcolor[HTML]{EFEFEF}   \smallsix{Output modification \tiny{(S3)}} \\
    \multicolumn{2}{l|}{\multirow{-3}{*}{Other Vulnerabilities}}  
        &\cite{tan2024unlink}  
        &\checkmark  & &
        & & & &\checkmark
        & \checkmark & &
        & & \checkmark &
        & \smallsix{Over-forgetting}
        & \smallsix{Improved unlearning algorithms \tiny{(S1)}}\\ 
\bottomrule

\end{tabular}
}
\caption{Summary of threats in machine unlearning systems. \textbf{Attack Roles}: Data contributors (R1), Requesting users (R2), Accessible users (R3); \textbf{Attack Goals}: Untargeted (G1), Targeted (G2), Privacy leakage (G3), Others (G4); \textbf{Adversarial Knowledge}: White-box (K1), Grey-box (K2), Black-box (K3); \textbf{Attack Phases}: Training phase (P1), Unlearning phase (P2), Post-unlearning phase (P3); \textbf{Defense Stages}: Pre-unlearning stage (S1), In-unlearning stage (S2), Post-unlearning stage (S3).}
\label{tab:summary}
\end{table*}

\subsection{Summary}
So far, we have explored threats inherent to machine unlearning systems and presented a taxonomy that includes:
\begin{itemize}
    \item Information leakage
    \item Malicious unlearning
    \item Other vulnerabilities
\end{itemize}

Additionally, we have conducted an in-depth analysis of their methodologies, guided by the threat models. A summary of threats in machine unlearning is provided in Table \ref{tab:summary}. Furthermore, in discussing defense strategies against these threats and attacks, a detailed review reveals that these strategies can be categorized based on the stage at which the defense is implemented. The classifications are as follows:

\begin{itemize}
    \item Pre-unlearning: at this stage, defense methods usually focus on detecting malicious requests prior to initiating the unlearning process, or implementing regulations to govern how unlearning is conducted.
    \item In-unlearning: at this stage, defense strategies often focus on monitoring changes in the model to potentially halt the unlearning process if anomalies are detected.
    \item Post-unlearning: at this stage, defense methods here generally aim at protecting information leaked from unlearned models, or recovering the model to its state before the attack, for example, by adding differential perturbations to the unlearned models, rolling back the model, or removing backdoors for model recovery, leveraging unlearning techniques (see Section \ref{sec:unlearning_as_defenses} for more details).
\end{itemize}

Note that defense strategies during the in-unlearning and post-unlearning stages are highly flexible, allowing for the use of diverse algorithms as drop-in solutions to mitigate both direct and preconditioned unlearning attacks. Conversely, the defensive approaches used in the pre-unlearning stage vary significantly, tailored to address specific threats and attacks.

\section{Defense through Unlearning}
\label{sec:unlearning_as_defenses}

As previously discussed, beyond facilitating knowledge removal, unlearning can also act as a defensive mechanism, collectively enhancing AI safety. Therefore, we dedicate this section to a focused discussion on unlearning as a form of defense. Specifically, we will explore two primary targets of unlearning in this defensive context, including model recovery and value alignment.

\subsection{Model Recovery}

Data poisoning attacks on machine learning models have been extensively studied over the years, revealing that both untargeted and targeted attacks can be effectively detected either during or after the training process\cite{li2020invisible,li2021hidden,zhang2022poison,han2023backdooring,zhang2023backdooring,wang2024transtroj,xu2021explainability}. However, these detection methods typically need substantial model updates to make confident decisions. As a result, by the time attacks are identified, the poisoned data may have already impacted the ML model \cite{cao2023fedrecover,ma2023beatrix,li2023ntd}. This scenario underscores the necessity for recovering an accurate model from a poisoned one that has been compromised after the detection of such attacks.

In this context, machine unlearning presents itself as an intuitive approach for model recovery, enabling the targeted removal of identified poisoned data. For example, \cite{wang2019neural} demonstrates backdoor removal by inverting the trigger to retrain infected ML models, a technique also utilized in \cite{guo2019tabor}. A more complex strategy for eliminating backdoors is detailed in \cite{liu2022backdoor}, where the process begins with identifying the trigger pattern based on which then conducting unlearning via gradient ascent. \cite{goel2024corrective} focuses on model recovery through unlearning, based solely on limited access to poisoned data. In \cite{liu2024model}, sparsity-aware unlearning is executed by initially pruning the model, followed by the unlearning process. Diverging from these approaches, \cite{zeng2021adversarial} and \cite{wei2024shared} explore the effects of unlearning on adversarial and backdoor attacks, each proposing different methods to balance trade-offs. The former focuses on unlearning universal adversarial perturbations, while the latter targets the unlearning of shared adversarial examples. The concept of model recovery through unlearning has also been explored in various other settings, including multimodal environments as seen in \cite{bansal2023cleanclip}, and federated learning scenarios, as discussed in \cite{cao2023fedrecover,jiang2024towards}.

\subsection{Value Alignment}

The concept of value alignment plays a pivotal role in enhancing AI safety, underscoring the importance of aligning machine learning models with human values and ethical standards. Given its significance, a wide array of research has been developed to leverage machine unlearning techniques as a strategic approach to achieve value alignment. This focus aims to mitigate risks and ensure that AI systems operate in a manner consistent with societal norms and individual preferences, highlighting unlearning as a key tool in the pursuit of safer AI integration.\\
\indent In these efforts, unlearning is applied in various innovative ways to align AI operations with ethical and legal standards. For instance, attribute unlearning \cite{li2023making} is used to remove sensitive attributes for privacy compliance. Similarly, unlearning with corresponding detection methods are applied to LLMs to remove behaviors and content that are deemed inappropriate or should be avoided, effectively eliminating illegal, poor quality, copyrighted, or harmful content \cite{liu2024rethinking,xue2024erase,yao2023large,li2024digger,liu2024towards,wang2023mitigating,guo2023fast,bano2023federated}. In a separate vein, the work outlined by Isonuma et al. \cite{isonuma2024unlearning} introduces a distinctive application of unlearning principles aimed at identifying the training datasets that have the most profound influence on generating harmful content. Through a process of systematically unlearning each dataset via gradient ascent, the repercussions on the model's propensity to generate harmful content can be meticulously evaluated post-unlearning. 

\section{Evaluating Unlearning through Attacks}
\label{sec:attacks_for_evaluation}

Much like a coin has two sides, attacks on unlearning systems can be both a threat and an effective tool for evaluation. This dual role enables the auditing of privacy leakage, the assessment of model robustness, and the provision of proof of unlearning. In this section, we provide an overview of methods for evaluating unlearning through attacks, highlighting how these aspects of attacks can deepen our understanding and enhance the effectiveness of unlearning systems.

\subsection{Audit of Privacy Leakage} 

As outlined in Section \ref{sec:information_leakage}, machine unlearning systems are susceptible to threats of information leakage. Attackers can exploit discrepancies between the trained and unlearned models, or inherent vulnerabilities within the unlearning system, to initiate inference attacks \cite{chen2021machine,lu2022label,gao2022deletion,hu2023eraser}, or to reconstruct the unlearned data \cite{chourasia2023forget,schwinn2024soft}.

Therefore, these attacks can function as instruments to audit privacy leakage from the unlearning system after the unlearning process, representing common methodologies in privacy leakage assessment \cite{sun2023not,hu2023veridip}. Metrics such as the Area Under the Curve (AUC) and Attack Success Rate (ASR) are widely adopted to quantify the effectiveness of privacy attacks, thereby indicating the level of privacy leakage \cite{chen2021machine,lu2022label,gao2022deletion,hu2023eraser}. Additionally, tracing the unlearning performance relative to the leakage of specific data with certain features can help measure overall privacy leakage \cite{wang2024selective}. Moreover, privacy degradation metrics, which compare against classical attacks to ML models, can quantify additional privacy leakage beyond what is typical with a single model \cite{chen2021machine}.

\subsection{Assessment of Model Robustness}

Similarly, malicious unlearning can assess the robustness of the unlearned model. The foundational principle of this assessment approach is akin to that used in adversarial attacks, as detailed in works such as \cite{cao2023stylefool,cao2024logostylefool,liu2019ppgan,cao2024localstylefool}. For example, \cite{hu2024duty} describes a method of malicious unlearning via over-unlearning, where perturbed data are submitted in the unlearning request to alter the classifier's decision boundary. This strategy of pushing the decision boundary allows for evaluating the impact of unlearned data points near the boundary on model performance, as well as the effect of data points that challenge the model's prediction capabilities. Another example occurs in the context of LLMs\cite{schwinn2024soft}, where attackers insert adversarial embeddings into the prompts' embeddings, aiming to extract supposedly unlearned knowledge from the model. Such attacks provide a means to gauge the model's resilience against malicious input embeddings. On a different note, \cite{zhang2022prompt} and \cite{che2023fast} treat data removal as an adversarial perturbation to the overall training dataset. In this case, the concept of randomized smoothing \cite{cohen2019certified} can be adopted to measure the model robustness against unlearned data \cite{zhang2022prompt} or unlearned users in a federated setting \cite{che2023fast}, hence establishing a certified budget for data removals but retains the model robustness.

\subsection{Proof of Unlearning}

As mentioned previously, membership inference attacks serve to identify information leakage within unlearning systems, while backdoor attacks aim to implant backdoors in models, leading to the models incorrectly classifying certain inputs when specific triggers are present. Viewing from a different angle, in a machine unlearning setting, if unlearned data show no membership in the model post-unlearning, or if backdoors within the unlearned data are undetectable, it is logical to conclude that the unlearned data has been effectively removed from the model with high probability. Based on this rationale, many studies on unlearning methodologies have employed the standard form or variants of inference attacks \cite{chundawat2023zero,ma2022learn,wang2022federated,liu2021federaser,liu2023breaking,alag2023ema,ye2023reinforcement} and backdoor attacks \cite{wu2201federated,sommer2020towards,guo2023verifying,pawelczyk2023context,halimi2022federated,li2023federated} as tools serving as proof of unlearning. Some other studies utilize adversarial attacks to detect the insufficiency of unlearning procedures \cite{zhang2023generate,goel2022towards,zeng2021adversarial,wei2024shared}. 
This list represents only a selection of studies, as the use of attacks to demonstrate unlearning has been thoroughly reviewed in prior literature. For those seeking more in-depth information on this field, we recommend referring to \cite{xuh2023machine,qu2023learn,shaik2023exploring,lynch2024eight,thaker2024guardrail} for comprehensive details.

\begin{figure}[t]
\centering
		\centering
\includegraphics[width=0.9\linewidth]{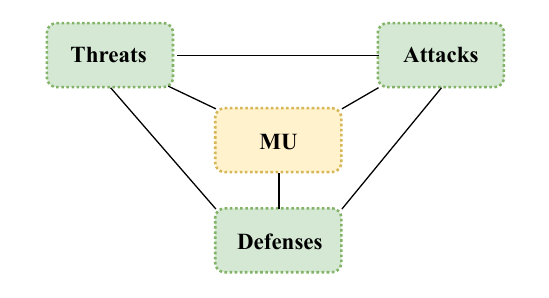}
		\caption{Threats, attacks, and defenses in machine unlearning.}
		\label{fig:relatinship}
\end{figure}

\section{Summary and Discussion}
\label{sec:summary}

So far, we have reviewed the related works on threats in MU systems, along with an analysis of various attack types, defense mechanisms, and their interactions in different roles. To clarify these relationships, as illustrated in Figure \ref{fig:relatinship}, this section offers a summary and discussion.

\textbf{Machine unlearning.} MU presents both threats and opportunities for defense: (i) the process of machine unlearning and the handling of unlearning requests introduce threats such as information leakage and the potential for malicious unlearning attacks, and (ii) unlearning can be used as a defensive strategy for model recovery and remove undesirable knowledge for value alignment.

\textbf{Threats.} Threats arise from the unlearning process and the workflow of MU systems, creating opportunities for attacks. Therefore, corresponding defensive strategies must be explored and developed.

\textbf{Attacks.} Attacks serve two key purposes: (i) to explore potential threats in MU systems, and (ii) to evaluate the performance of unlearning systems, including aspects such as privacy leakage, robustness, and effectiveness.

\textbf{Defenses.} As mentioned earlier, (i) MU can serve as a defense in certain scenarios, and (ii) defenses against specific threats in MU systems still need to be thoroughly explored.

\section{Challenges and Promising Directions}
\label{sec:discussions_and_promising_directions}

In earlier sections, we provided a detailed overview of threats, attacks, and defenses within machine unlearning systems. Yet, as unlearning methodologies continue to advance and their application expands, a range of new challenges and open problems emerge, necessitating further exploration. Consequently, this section aims to expand our discussion on these emerging challenges, suggesting potential directions for future research that could significantly improve the safety and reliability of machine unlearning systems.

\subsection{Defenses against Malicious Unlearning}

Note that one of the main threats to machine unlearning systems is the negative impact on unlearned models caused by malicious unlearning (see Section \ref{sec:malicous_unlearning} for more details). This threat manifests as either degraded model performance or the introduction of backdoors, involving the submission of malicious unlearning requests. While direct malicious unlearning requests can often be easily detected by checking if the requested data for removal is part of the training dataset \cite{hu2024duty}, identifying such requests during preconditioned unlearning attacks presents a nontrivial challenge. This is because the data requested for removal in these attacks are actually part of the training dataset and usually are mitigation data without adversarial features. Therefore, developing robust detection mechanisms for identifying these subtle and sophisticated unlearning requests becomes essential in safeguarding the integrity of machine unlearning systems.

\subsection{Federated Unlearning}

As highlighted in \cite{liu2023survey,jeong2024sok}, federated learning's distinct features pose specific challenges to machine unlearning methodologies, largely due to knowledge diffusion through aggregation and data isolation aimed at protecting user data privacy. Consequently, threats in a federated unlearning context are inherently more complex. Additionally, the involvement of multiple users in federated unlearning means that attacks executed in a distributed fashion can be particularly stealthy and difficult to detect. However, this area remains largely unexplored, indicating a significant gap in current research.

In addition, within a federated setting, users' data are stored locally, prohibiting the transfer of local data to the server as it would violate the fundamental privacy guarantees of federated learning. Thus, understanding how unlearning requests are made, their specific format, and how these requests are managed by the server in a federated unlearning context is crucial. Studying these aspects is essential to developing an unlearning mechanism that adheres to both the privacy principles of federated learning and the compliance requirements of RTBF.

\subsection{Privacy Preservation}

Current machine unlearning systems operate under the assumption that the data designated for removal is accessible to the server responsible for the unlearning process. However, in an MLaaS setting, the model developer and service provider may be distinct entities \cite{hu2024duty}. In such scenarios, the service provider should not access the user's data, raising questions about how to protect the privacy of the unlearned data. This challenge echoes the privacy concerns highlighted in the federated learning context.

In general, one could employ Privacy-Enhancing Technologies (PETs) like homomorphic encryption, secure multi-party computation, or differential privacy to facilitate privacy-preserving machine learning \cite{MZ17,liu2022efficient,liu2022privacy,lam2024efficient,liu2024dynamic,liu2023long,wagh2019securenn}. Nevertheless, given that machine unlearning methodologies significantly diverge from traditional machine learning practices and introduce new privacy concerns, achieving privacy-preserving machine unlearning presents new challenges. Although there has been some investigation into privacy-preserving machine unlearning \cite{liu2024guaranteeing,liu2024privacy}, the trade-offs introduced by these technologies between privacy protection, unlearning efficiency, and model performance in the context of MU remain largely unexplored.

\subsection{Unlearning for Large Models}

Investigating the unlearning for large models is an emerging and vital area of research, crucial for enhancing the security and robustness of large model systems. This exploration faces several challenges, including:
\begin{itemize}
    \item The inefficiency of the unlearning for large models, which necessitates innovative solutions to improve its effectiveness
    \cite{liu2024machine}.
    \item The difficulty in verifying unlearning effectiveness in large models, as evidenced by \cite{yao2023large}, suggests that methods like MIA used for proving unlearning are less effective in larger models.
    \item The inherent non-explainability of large models complicates understanding the unlearning process. For instance, \cite{schwinn2024soft} demonstrates that unlearned information can still be retrieved through embedding attacks after unlearning.
    \item The exploration of threats and attacks specific to unlearning in large models is still lacking.
\end{itemize}

Given the widespread deployment of large models in numerous applications \cite{pan2024unifying,chen2023separate,kasneci2023chatgpt,chen2023dipping,thirunavukarasu2023large,chen2022knowledge,deng2024masterkey}, tackling these challenges and intensifying research on threats, attacks, and defenses within large model unlearning systems is imperative.

\section{Conclusions}
\label{sec:conclusions}
This survey examines machine unlearning as a vital component of efforts towards achieving safe AI, concentrating on threats and attacks against these systems, alongside the defensive strategies employed to counteract them. It provides a detailed analysis of methodologies and creates a comprehensive taxonomy based on their threat models. Furthermore, the survey investigates how unlearning can act as a defense in different scenarios and how attacks can serve as effective tools for testing and evaluating unlearning systems. Lastly, this work identifies challenges and outlines future directions for improving the safety, reliability, and privacy compliance of machine unlearning, stressing the critical need for continued exploration in this area.


\bibliographystyle{IEEEtran}
\bibliography{mybibliography}

\end{document}